\documentclass[twocolumn]{aastex63}

\usepackage{upgreek}

\newcommand\kms{km~s$^{-1}$}
\newcommand\msun{M$_\odot$}
\newcommand\rsun{R$_\odot$}

\received{XXXX XX, 2021}
\revised{XXXX XX, 2021}
\accepted{XXXX XX, 2021}
\submitjournal{}

\shorttitle{SN~2021csp}
\shortauthors{Fraser et al.}

\begin{document}


\title[SN~2021csp]{SN~2021csp - the explosion of a stripped envelope star within a H and He-poor circumstellar medium}

\correspondingauthor{Morgan Fraser}
\email{morgan.fraser@ucd.ie}
\author[0000-0003-2191-1674]{Morgan Fraser}
\affiliation{School of Physics, University College Dublin, Belfield, Dublin 4, Ireland}

\author[0000-0002-5571-1833]{Maximilian D. Stritzinger} 
\affiliation{Department of Physics and Astronomy, Aarhus University, Ny Munkegade 120, DK-8000 Aarhus C, Denmark}

\author[0000-0003-1325-6235]{Se\'an J. Brennan} 
\affiliation{School of Physics, University College Dublin, Belfield, Dublin 4, Ireland}

\author[0000-0002-7259-4624]{Andrea Pastorello}
\affiliation{INAF -- Osservatorio Astronomico di Padova, Vicolo dell’Osservatorio  5, I-35122 Padova, Italy}

\author[0000-0002-7714-493X]{Yongzhi Cai} 
\affiliation{Physics Department and Tsinghua Center for Astrophysics (THCA), Tsinghua University, Beijing 100084, China}

\author[0000-0001-6806-0673]{Anthony L. Piro} 
\affiliation{The Observatories of the Carnegie Institution for Science, 813 Santa Barbara St., Pasadena, CA 91101, USA}

\author[0000-0002-5221-7557]{Chris Ashall} 
\affiliation{Institute for Astronomy, University of Hawai'i, 2680 Woodlawn Drive, Honolulu, HI 96822, USA} 

\author[0000-0001-6272-5507]{Peter Brown} 
\affiliation{George P. and Cynthia Woods Mitchell Institute for Fundamental Physics and Astronomy, Department of Physics and Astronomy, Texas A \& M University, College Station, TX 77843, USA} 

\author[0000-0003-4625-6629]{Christopher R. Burns} 
\affiliation{The Observatories of the Carnegie Institution for Science, 813 Santa Barbara St., Pasadena, CA 91101, USA}

\author[0000-0002-1381-9125]{Nancy Elias-Rosa} 
\affiliation{INAF -- Osservatorio Astronomico di Padova, Vicolo dell’Osservatorio  5, I-35122 Padova, Italy}
\affiliation{Institute of Space Sciences (ICE, CSIC), Campus UAB, Carrer de Can Magrans s/n, E-08193 Barcelona, Spain} 

\author[0000-0003-3460-0103]{Alexei V. Filippenko}
\altaffiliation{Miller Senior Fellow}
\affiliation{Department of Astronomy, University of California, Berkeley, CA 94720-3411, USA}
\affiliation{Miller Institute for Basic Research in Science, University of California, Berkeley, CA 94720, USA}

\author[0000-0002-1296-6887]{L. Galbany} 
\affiliation{Institute of Space Sciences (ICE, CSIC), Campus UAB, Carrer de Can Magrans s/n, E-08193 Barcelona, Spain} 

\author[0000-0003-1039-2928]{E. Y. Hsiao}
\affiliation{Department of Physics, Florida State University, 77 Chieftain Way, Tallahassee, FL, 32306, USA} 

\author[0000-0001-8738-6011]{Saurabh W. Jha}
\affiliation{Department of Physics and Astronomy, Rutgers the State University of New Jersey, 136 Frelinghuysen Road, Piscataway, NJ 08854 USA}

\author[0000-0001-5455-3653]{Rubina Kotak} 
\affiliation{Tuorla Observatory, Department of Physics and Astronomy, University of Turku, 20014, Turku, Finland}

\author[0000-0001-5221-0243]{Shane Moran}
\affiliation{Tuorla Observatory, Department of Physics and Astronomy, University of Turku, 20014, Turku, Finland}

\author[0000-0003-2535-3091]{Nidia Morrell}
\affiliation{Carnegie Observatories, Las Campanas Observatory, Casilla 601, La Serena, Chile} 

\author{Paolo Ochner}
\affiliation{Dipartimento di Fisica e Astronomia, Universit\`a di Padova, Vicolo Osservatorio 2, 35122, Padova, Italy}
\affiliation{INAF -- Osservatorio Astronomico di Padova, Vicolo dell’Osservatorio  5, I-35122 Padova, Italy}

\author[0000-0003-2734-0796]{M. M. Phillips} 
\affiliation{Carnegie Observatories, Las Campanas Observatory, Casilla 601, La Serena, Chile} 

\author[0000-0003-4254-2724]{Andrea Reguitti}
\affiliation{Departamento de Ciencias Fisicas, Universidad Andres Bello, Avda. Republica 252, Santiago, Chile}
\affiliation{Millennium Institute of Astrophysics (MAS), Nuncio Monse\~nor Sotero Sanz 100, Providencia, Santiago, Chile}
\affiliation{INAF -- Osservatorio Astronomico di Padova, Vicolo dell’Osservatorio  5, I-35122 Padova, Italy}

\author[0000-0003-4631-1149]{B.~J.~Shappee}
\affiliation{Institute for Astronomy, University of Hawai'i, 2680 Woodlawn Drive, Honolulu, HI 96822, USA}

\author[0000-0002-3697-2616]{Lina Tomasella} 
\affiliation{INAF -- Osservatorio Astronomico di Padova, Vicolo dell’Osservatorio  5, I-35122 Padova, Italy}

\author[0000-0002-2471-8442]{Michael Tucker}
\altaffiliation{DOE CSGF Fellow}
\affiliation{Institute for Astronomy, University of Hawai'i, 2680 Woodlawn Drive, Honolulu, HI 96822, USA}

\author{Xiaofeng Wang}
\affiliation{Physics Department and Tsinghua Center for Astrophysics (THCA), Tsinghua University, Beijing 100084, China}
\affiliation{Beijing Planetarium, Beijing Academy of Science and Technology, Beijing 100044, China} 

\author[0000-0002-8296-2590 ]{Ju-jia Zhang} 
\affiliation{Yunnan Observatories, Chinese Academy of Sciences, Kunming 650216, China}
\affiliation{Key Laboratory for the Structure and Evolution of Celestial Objects, Chinese Academy of Sciences, Kunming 650216, China}

\author[0000-0003-0227-3451]{J. P. Anderson} 
\affiliation{European Southern Observatory, Alonso de C\'ordova 3107, Casilla 19, Santiago, Chile}

\author[0000-0002-4843-345X]{Tyler Barna}
\affiliation{Department of Physics and Astronomy, Rutgers the State University of New Jersey, 136 Frelinghuysen Road, Piscataway, NJ 08854 USA}

\author[0000-0001-5393-1608]{E. Baron} 
\affiliation{Homer L. Dodge Department of Physics and Astronomy, University of Oklahoma, Rm 100, 440 W. Brooks, Norman, OK 73019-20615, USA}
\affiliation{Hamburger Sternwarte, Gojenbergsweg 112, 21029 Hamburg, Germany}

\author[0000-0002-3256-0016]{Stefano Benetti}
\affiliation{INAF -- Osservatorio Astronomico di Padova, Vicolo dell’Osservatorio  5, I-35122 Padova, Italy}

\author[0000-0002-6991-0550]{Melina C. Bersten} 
\affiliation{Facultad de Ciencias Astron\'omicas y Geof\'isicas, Universidad Nacional de La Plata, Paseo del Bosque S/N, B1900FWA, La Plata, Argentina}
\affiliation{Instituto de Astrof\'isica de La Plata (IALP), CCT-CONICET-UNLP. Paseo del Bosque S/N, B1900FWA, La Plata, Argentina}
\affiliation{Kavli Institute for the Physics and Mathematics of the Universe(WPI), The University of Tokyo, 5-1-5 Kashiwanoha, Kashiwa, Chiba 277-8583, Japan}

\author{Thomas G. Brink}
\affiliation{Department of Astronomy, University of California, Berkeley, CA 94720-3411, USA}

\author[0000-0002-9830-3880]{Yssavo Camacho-Neves}
\affiliation{Department of Physics and Astronomy, Rutgers the State University of New Jersey, 136 Frelinghuysen Road, Piscataway, NJ 08854 USA}

\author{Scott Davis} 
\affiliation{Department of Physics and Astronomy, University of California, 1 Shields Avenue, Davis, CA 95616-5270, USA}

\author[0000-0001-7519-133X]{Kyle G. Dettman}
\affiliation{Department of Physics and Astronomy, Rutgers, the State University of New Jersey, 136 Frelinghuysen Road, Piscataway, NJ 08854 USA}

\author[0000-0001-5247-1486]{Gaston Folatelli} 
\affiliation{Facultad de Ciencias Astron\'omicas y Geof\'isicas, Universidad Nacional de La Plata, Paseo del Bosque S/N, B1900FWA, La Plata, Argentina}
\affiliation{Instituto de Astrof\'isica de La Plata (IALP), CCT-CONICET-UNLP. Paseo del Bosque S/N, B1900FWA, La Plata, Argentina}
\affiliation{Kavli Institute for the Physics and Mathematics of the Universe (WPI), The University of Tokyo, 5-1-5 Kashiwanoha, Kashiwa, Chiba 277-8583, Japan}

\author[0000-0003-2375-2064]{Claudia P. Guti\'errez}
\affiliation{Finnish Centre for Astronomy with ESO (FINCA), FI-20014 University of Turku, Finland}
\affiliation{Tuorla Observatory, Department of Physics and Astronomy, University of Turku, 20014, Turku, Finland}

\author[0000-0002-4338-6586]{Peter H\"{o}flich} 
\affiliation{Department of Physics, Florida State University, 77 Chieftain Way, Tallahassee, FL, 32306, USA}

\author[0000-0001-9206-3460]{Thomas~W.-S.~Holoien} 
\altaffiliation{NHFP Einstein Fellow}
\affiliation{The Observatories of the Carnegie Institution for Science, 813 Santa Barbara St., Pasadena, CA 91101, USA}

\author[0000-0001-8257-3512]{Erkki Kankare}
\affiliation{Tuorla Observatory, Department of Physics and Astronomy, University of Turku, 20014, Turku, Finland}

\author[0000-0002-1132-1366]{Hanindyo Kuncarayakti} 
\affiliation{Tuorla Observatory, Department of Physics and Astronomy, University of Turku, 20014, Turku, Finland}
\affiliation{Finnish Centre for Astronomy with ESO (FINCA), FI-20014 University of Turku, Finland}

\author[0000-0003-3108-1328]{Lindsey Kwok} 
\affiliation{Department of Physics and Astronomy, Rutgers the State University of New Jersey, 136 Frelinghuysen Road, Piscataway, NJ 08854 USA}

\author{Sahana Kumar} 
\affiliation{Department of Physics, Florida State University, 77 Chieftain Way, Tallahassee, FL, 32306, USA} 

\author[0000-0002-3900-1452]{Jing Lu} 
\affiliation{Department of Physics, Florida State University, 77 Chieftain Way, Tallahassee, FL, 32306, USA} 

\author[0000-0001-6876-8284]{Paolo Mazzali}
\affiliation{Astrophysics Research Institute, Liverpool John Moores University, IC2, Liverpool Science Park, 146 Brownlow Hill, Liverpool L3 5RF, UK}
\affiliation{Max-Planck-Institut f{\"u}r Astrophysik, Karl-Schwarzschild Str. 1, D-85748 Garching, Germany}

\author[0000-0002-9301-5302]{Melissa Shahbandeh} 
\affiliation{Department of Physics, Florida State University, 77 Chieftain Way, Tallahassee, FL, 32306, USA} 

\author[0000-0002-8102-181X]{Nicholas B. Suntzeff}
\affiliation{George P. and Cynthia Woods Mitchell Institute for Fundamental Physics and Astronomy, Department of Physics and Astronomy, Texas A \& M University, College Station, TX 77843, USA}

\author{Stefan Taubenberger} 
\affiliation{Max-Planck-Institut f{\"u}r Astrophysik, Karl-Schwarzschild Str. 1, D-85748 Garching, Germany}

\author[0000-0002-1481-4676]{Samaporn Tinyanont} 
\affiliation{Department of Astronomy and Astrophysics, University of California, Santa Cruz, CA 95064, USA}

\author{Shengyu Yan} 
\affiliation{Physics Department and Tsinghua Center for Astrophysics (THCA), Tsinghua University, Beijing 100084, China}

\author{Yi Yang} 
\altaffiliation{Bengier-Winslow-Robertson Fellow}
\affiliation{Department of Astronomy, University of California, Berkeley, CA 94720-3411, USA}

\author{WeiKang Zheng} 
\affiliation{Department of Astronomy, University of California, Berkeley, CA 94720-3411, USA}



\begin{abstract}

We present observations of SN~2021csp, a unique supernova (SN) which displays evidence for interaction with H- and He- poor circumstellar material (CSM) at early times. Using high-cadence spectroscopy taken over the first week after explosion, we show that the spectra of SN~2021csp are dominated by \ion{C}{3} lines with a velocity of 1800~\kms. We associate this emission with CSM lost by the progenitor prior to explosion. Subsequently, the SN displays narrow He lines before metamorphosing into a broad-lined Type Ic SN. We model the bolometric light curve of SN~2021csp, and show that it is consistent with the energetic ($4\times10^{51}$~erg) explosion of a stripped star, producing 0.4~\msun\ of $^{56}$Ni within a $\sim 1$~\msun\ shell of CSM extending out to 400 \rsun.

\end{abstract}



\section{Introduction}
\label{sec:intro}

Solar-metallicity stars born with masses above $\sim 30$~\msun\ are expected to lose their entire H (or even He envelope) before exploding as stripped-envelope (SE) core-collapse supernovae (SNe). This mass loss can be a result of strong stellar winds \citep{Conti75}, as have been observed in Wolf-Rayet stars (although we note that single-star models struggle to remove the entire He envelope). Alternatively, mass transfer onto a binary companion can strip massive stars \citep[e.g.,][]{Podsi92}; this channel can also produce SESNe from lower-mass (10--30~\msun) progenitors.

This progenitor paradigm for SESNe is supported by a number of lines of direct and indirect evidence, including nebular-phase spectroscopy \citep[e.g.,][]{Kuncarayakti15,Fremling16,Teffs21},  direct detections of SESN progenitors \citep[e.g.,][]{Eldridge13, Groh13, Cao13, Eldridge15, Folatelli16, Kilpatrick18, Xiang19, Kilpatrick21}, and estimates of SESN ejecta masses \citep[e.g.,][]{Drout11,Lyman2016,Prentice16,taddia2018}.

A small number of SESNe have been observed to have relatively narrow ($\lesssim 1000$~\kms) He lines in their spectra \citep{Pastorello2007,Foley2007}. These events are known as Type Ibn SNe, and the narrow lines are thought to form in a region of H-poor circumstellar material (CSM) that surrounded the progenitor prior to explosion. Besides SNe~Ibn, a number of SESNe have shown evidence for interaction with {\it H-rich} CSM at late times in the form of strong, relatively narrow Balmer lines \citep{Milisavljevic2015,Kuncarayakti18}, or a rebrightening powered by CSM interaction \citep{Sollerman2020}. Along with normal SESNe, some members of the energetic subclasses of broad-lined and superluminous SESNe have also developed relatively narrow H lines at late phases \citep{Chen2018, Yan2015}. The peculiar SE SN~2010mb displayed evidence for interaction with H-free CSM at late times \citep{BenAmi14}.

More recently, a number of authors have reported evidence for CSM interaction in some SESNe at early phases. \cite{Ho19} found a pre-explosion outburst prior to the SE SN~2018ge, and suggested that the fast rise and high temperatures at early times were consistent with shock breakout from a dense CSM. Highly ionized ``flash'' features that likely arise from the pre-SN wind of the progenitor have also been seen in two SNe~Ibn \citep{Shivvers16,Gangopadhyay20}.

Together, these events paint a picture of a heterogeneous population of SESN progenitors, with a wide range of progenitor masses, mass-loss rates, and CSM configurations \citep{Fraser2020}. However, many questions remain unanswered, including the composition and structure of the CSM, the role of binarity, and whether eruptive mass loss has a significant role in removing the envelope of SESN progenitors.

Among  SESNe, the subclass of broad-lined (BL) SNe~Ic (hereafter SNe~Ic-BL) has attracted considerable interest thanks to their association with gamma-ray bursts (GRBs), first seen in the case of GRB~980425 / SN~1998bw \citep{galama1998,iwamoto1998}. 
By definition, the optical spectra of SNe~Ic-BL lack  signatures of H and He,
and display high ejecta velocities on the order of 20,000~\kms, resulting in broad spectral features.
As mentioned previously, a handful of SNe~Ic-BL have been connected to H-rich CSM at late times \citep[e.g., SN~2017ens;][]{Chen2018} or show signs of early interaction with CSM \cite[iPTF16asu and  SN~2018gep;][]{Ho19,Whitesides17,Wang19,Pritchard20}.

In this paper, we report observations of the recently discovered SN~2021csp, which uniquely shows evidence for interaction with an H {\it and} He-poor CSM.
SN~2021csp was first discovered by \citet{21cspdiscovery} as ZTF21aakilyd with J2000 coordinates $\alpha = 14^h 26^m 22^s.11$, $\delta = +05^\circ 51' 33''.2$ on 2021 Feb. 11.5 (UT dates are used throughout this paper) at $g = 18.1$~mag during the course of the Zwicky Transient Survey (ZTF; \citealt{Bellm19}). 
A nondetection by ZTF two days prior to the discovery with a limit of $g \approx 20.3$~mag suggested that the transient was fast rising.
A nearly simultaneous independent discovery of SN~2021csp (as ATLAS21ewe) on Feb 11.6 was made by the ATLAS survey (Asteroid Terrestrial-impact Last Alert System; \citealt{Tonry18}).

The first spectroscopic classification of SN~2021csp was made by \citet{perley2021}, based on a spectrum taken with the Liverpool Telescope (LT) + SPectrograph for the Rapid Acquisition of Transients (SPRAT) less than two days after discovery. \citeauthor{perley2021} observed neither H nor He lines, but reported the presence of P~Cygni features they attributed to \ion{C}{3}, \ion{O}{3}, and \ion{C}{4}. 
The presence of conspicuous, narrow C features led \citet{GalYamIcn} to suggest SN~2021csp to be a member of a newly designated class of ``SNe~Icn.''

Owing to its unusual spectral properties and rapid photometric evolution, the NUTS2\footnote{\url{https://nuts.sn.ie/}} (Nordic optical telescope Un-biased Transient Survey; \citealt{holmbo2019})
and POISE (Precision Observations of Infant Supernova Explosions;  \citealp{Burns2021}) collaborations began a high-cadence follow-up campaign for SN~2021csp.
POISE is a new project which aims to exploit high-precision, rapid-cadence, multiwavelength observations shortly after explosions with resources at Las Campanas Observatory similar to those utilized by the Carnegie Supernova Project-II (CSP-II; \citealp{Phillips19}; \citealp{Hsiao19}). The follow-up targets are strictly limited to those with deep ($>20$~mag) nondetection limits within 2~d prior to discovery.

In Sec.~\ref{sec:obs_data} we describe our observations and data reduction. The spectroscopic and photometric evolution of SN~2021csp are characterized in Sec.~\ref{sec:spec_evol} and \ref{sec:phot_evol}, respectively. In Sec.~\ref{sect:bolom_model}, we determine the bolometric light curve of SN~2021csp, and use simple modeling to estimate the ejecta mass and other physical parameters. Finally, in Sec.~\ref{sect:nature}, we discuss possible explanations for SN~2021csp. Throughout the following we adopt the earliest recovered detection of SN~2021csp (MJD 59255.47; see Sec.~\ref{sect:surveys}) as our reference epoch.
We take the distance modulus of SN~2021csp to be $\mu=37.77^{+0.04}_{-0.05}$~mag [from the measured redshift $z=0.083$ (Sec.~\ref{sec:spec_evol}), and adopting H$_0=74.0\pm1.42$~\kms~Mpc$^{-1}$, $\Omega_M=0.31$, $\Omega_\lambda=0.69$; \citealp{Riess2019}].
We assume the foreground reddening toward SN~2021csp to be $E(B-V)=0.028$~mag \citep{Schlafly_dust}, and consider the host-galaxy extinction to be negligible (Sec.~\ref{sec:spec_evol}).

\begin{figure}
\includegraphics[width=\columnwidth]{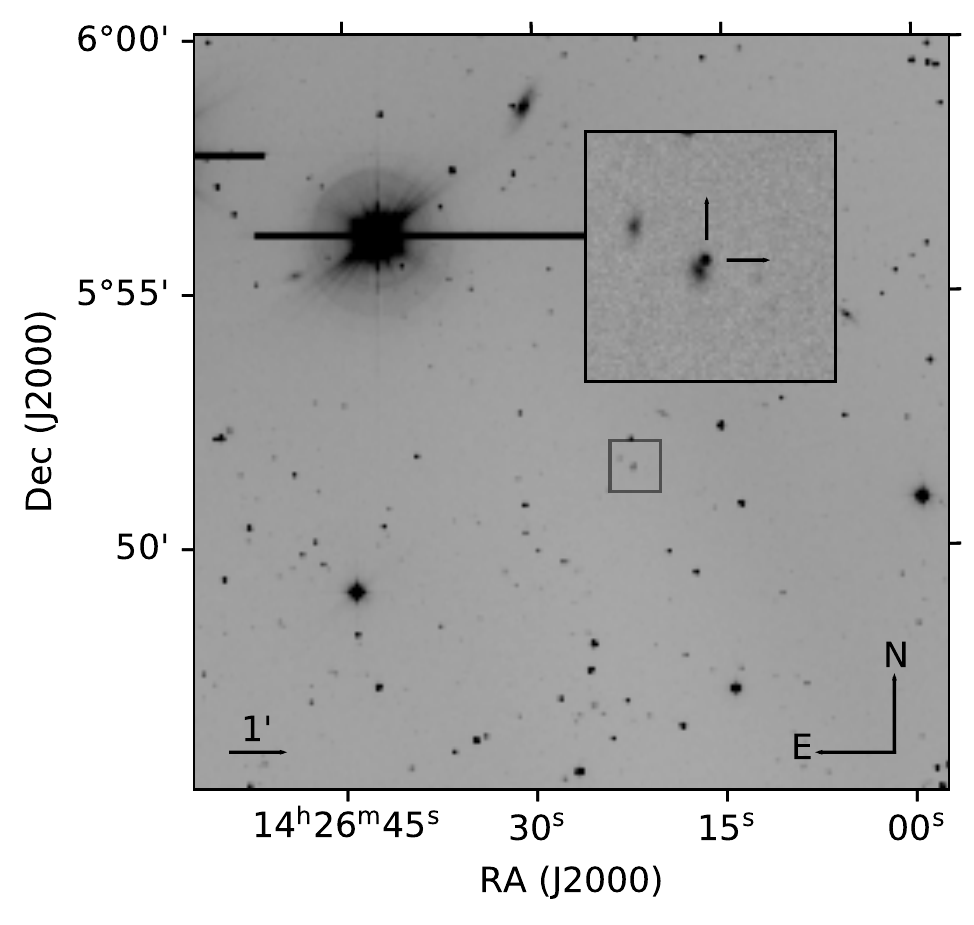}
\caption{Finding chart for SN~2021csp based on an $r$-band image taken with the 1~m Swope telescope. The inset shows the region around SN~2021csp in more detail.
\label{fig:finder}}
\end{figure}

\section{Observational data} \label{sec:obs_data}

\subsection{Spectroscopy} 
\label{sec:spec_data}

We obtained a comprehensive visual-wavelength spectroscopic time series of SN~2021csp using several different telescopes and instruments, including the 1.82~m Copernico Telcscope + Asiago Faint Object Spectrograph and Camera (AFOSC), the Nordic Optical Telescope +  Alhambra Faint Object Spectrograph and Camera (ALFOSC), the Gran Telescopio Canarias + Optical System for Imaging and low-Intermediate-Resolution Integrated Spectroscopy (OSIRIS; \citealp{Cepa2010}), the Magellan-Baade 6.5~m telescope + Inamori-Magellan Areal Camera and Spectrograph (IMACS; \citealp{Dressler2006}), the Lick Shane telescope + Kast spectrograph \citep{Miller93}, the  Telescopio Nazionale Galileo + Device Optimized for the LOw RESolution (DOLORES), the Lijiang 2.4~m Telescope + Yunnan Faint Object Spectrograph and Camera (YFOSC; \citealp{Wang19yfosc}), the Keck-I 10~m telescope + Low Resolution Imaging Spectrometer (LRIS; \citealp{Oke95}), and the 
University of Hawaii 2.2~m telescope + SuperNova Integral Field Spectrograph (SNIFS; \citealp{Lantz04}).
The temporal series consists of 21 epochs  covering the evolution of SN~2021csp from +2.6~d to +51.8~d.
In Table~\ref{tab:spec} we present the log of spectroscopic observations and list for each spectrum the facility, date of observations, resolution, and spectral range. 

\begin{figure*}
\includegraphics[width=0.85\textwidth]{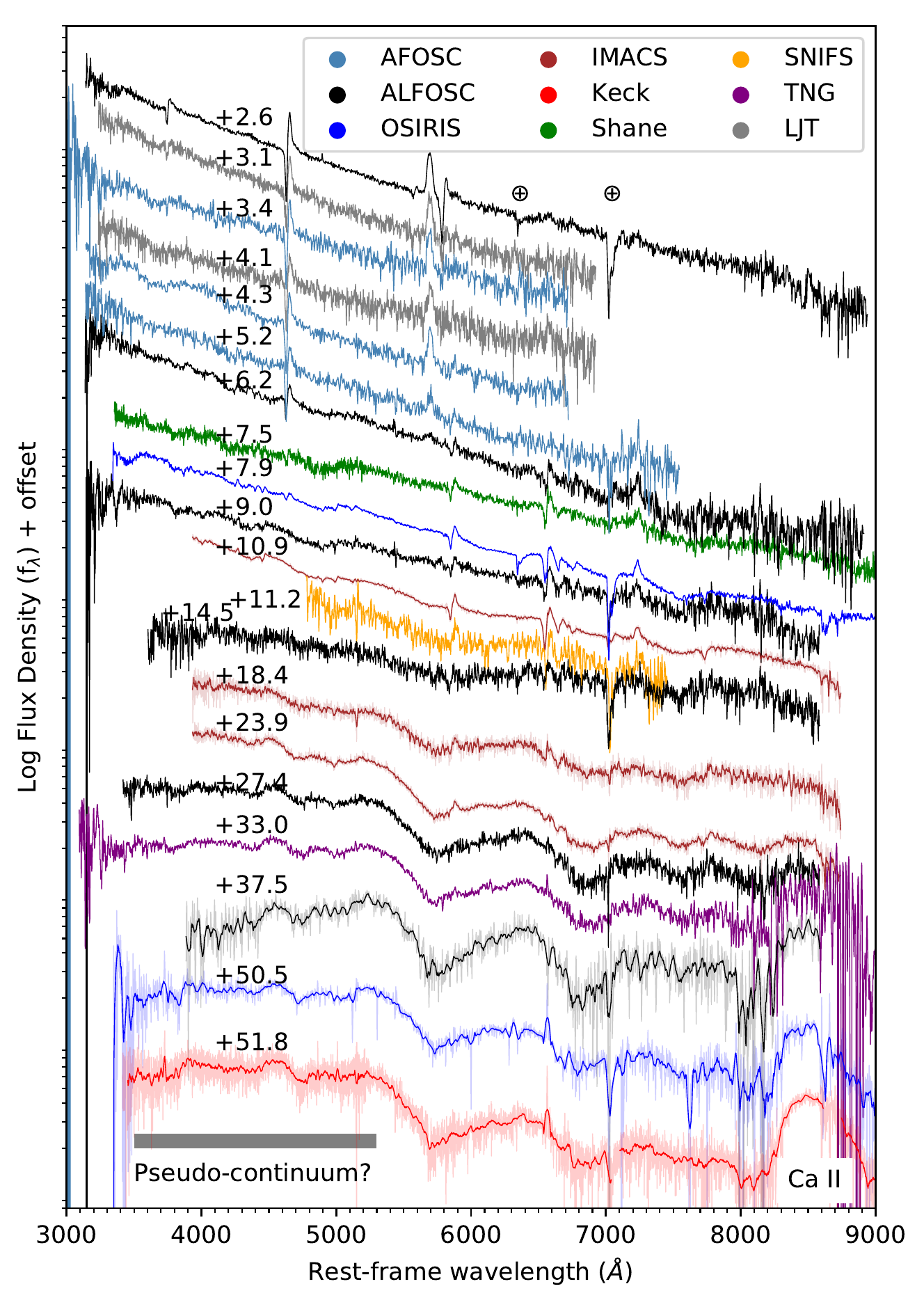}
\caption{Spectroscopic sequence of SN~2021csp, extending from $+$2.6~d to $+$51.8~d post discovery. Some low-S/N spectra have been omitted for clarity; wavelength regions with low S/N have also been trimmed. Spectra have been corrected for reddening. In some cases spectra with lower S/N have been smoothed with a  Savitzky-Golay filter. Where smoothing has been applied, the original spectrum is plotted as a light line, and the smoothed spectrum is a heavy line.}
\label{fig:spec_sequence}
\end{figure*}

We reduced all long-slit spectroscopic data following standard techniques. Two-dimensional spectra were overscan and bias-subtracted, before being divided by a normalized flat field. One-dimensional spectra of SN~2021csp were optimally extracted and wavelength-calibrated using contemporaneous spectra of arc lamps. The spectra were flux-calibrated using sensitivity curves derived from observations of spectrophotometric standard stars. In some cases, a smooth-continuum standard star was observed during the same night, in which case the strong telluric absorptions in the spectra of SN~2021csp were removed. For ALFOSC and AFOSC the {\sc alfoscgui} and {\sc afoscgui} pipelines\footnote{FOSCGUI is a graphical user interface aimed at extracting SN spectroscopy and photometry obtained with FOSC-like instruments. It was developed by E. Cappellaro. A package description can be found at \url{http://sngroup.oapd.inaf.it/foscgui.html.}} were used to reduce the data; for LRIS this was done with LPIPE \citep{LPIPE}. For all other instruments these steps were done within {\sc iraf}\footnote{IRAF is distributed by the National Optical Astronomy Observatories, which are operated by the Association of Universities for Research in Astronomy, Inc., under cooperative agreement with the National Science Foundation (NSF).}.

A single R-channel spectrum of SN~2021csp was taken with SNIFS \citep{Aldering02,Lantz04} at +11.2~d after discovery covering 3400--8080~\AA. Basic reduction details are described by \cite{Bacon01}; a custom Python-based pipeline reconstructs the spectral cube, extracts the spectrum using a low-order polynomial trace, and applies spectrophotometric and telluric corrections using a library of standard-star observations. 

In addition to the optical data, we obtained two epochs of near-infrared (NIR) spectroscopy of SN~2021csp with the Keck-II telescope equipped with the Near-Infrared Echellette Spectrometer (NIRES; \citealp{Wilson04}). For each NIR spectrum, one set of ABBA exposures was taken, where the duration of each individual exposure was 300~s. This led to a total time of 1200~s on target. The data were  reduced following standard procedures described in the  \texttt{IDL} package Spextools version 5.0.2 for NIRES \citep{spextool}. The extracted one-dimensional spectrum was flux-calibrated and corrected for telluric features  with \texttt{Xtellcorr} version 5.0.2 for NIRES, making use of observations of an A0V standard star \citep{spextool}.

\subsection{Imaging survey data}
\label{sect:surveys}

SN~2021csp was observed by a number of wide-field imaging surveys, namely ZTF \citep{Bellm19}, ATLAS \citep{Tonry18}, and the All-Sky Automated Survey for Supernovae (ASAS-SN; \citealt{Shappee14,Kochanek2017}).

ATLAS is a wide-field survey which observes the entire night sky accessible from Hawai'i in two filters, $c$ and $o$ (``cyan'' and ``orange,'' broadly similar to $g+r$ and $r+i$, respectively). 
The ATLAS forced-photometry server \citep{Smith20} was used to perform  photometry at the location of SN~2021csp in host-galaxy template-subtracted images. As ATLAS typically takes four consecutive exposures of any given field,  a weighted mean from the individual measured fluxes was computed at each epoch, which was then converted to an AB magnitude. This photometry is reported in Table~\ref{tab:phot_sloan}.

ZTF obtains $g$- and $r$-band images with a typical cadence of 2--3~d. Photometry of transients is measured on template-subtracted survey images. The ZTF photometry of SN~2021csp was accessed using  the LASAIR event broker \citep{Smith19}.

We also present $g$-band imaging of SN~2021csp obtained by  ASAS-SN, which observes the entire visible sky (weather dependent) nightly to a limit of $g \approx 18.5$~mag. We recovered our earliest detection of SN~2021csp at $g=18.8\pm0.03$~mag in ASAS-SN images taken on Feb. 10.5 (MJD 59255.47).

\subsection{Targeted follow-up imaging}

We obtained follow-up imaging of SN~2021csp using a number of facilities equipped with  Sloan ($ugriz$) and Johnson-Cousins ($BV$) filters. These  include the Las Campanas Observatory 1~m Swope Telescope + CCD ($uBVgri$), the 2.0~m Liverpool Telescope + IO:O ($uBVgriz$), the 1.82~m Copernico Telescope + AFOSC ($uBVg$), the 0.67/0.92~m Asiago Schmidt + CCD ($BVgri$), and the 2.56~m Nordic Optical Telescope (NOT) + ALFOSC ($uVgriz$).

Images were reduced following standard techniques including bias and overscan subtraction, flat-fielding, and (when appropriate) fringing corrections.
The Swope images were obtained by POISE
and reduced using the CSP \citep{krisciunas2017,Phillips19}) imaging-reduction pipeline, while
the LT images were reduced with the IO:O\footnote{\url{https://telescope.livjm.ac.uk/TelInst/Pipelines/\#ioo}} pipeline.
Finally, in the case of the Asiago Schmidt, Copernico, and NOT data, the {\sc pyraf}-based {\sc schmitgui}, {\sc afoscgui} and {\sc alfoscgui} pipelines were used.\footnote{\url{http://sngroup.oapd.inaf.it/foscgui.html}}

Point-spread-function (PSF) photometry was performed on all images using the {\sc AutoPhOT} pipeline \citep{Brennan2016}, and calibrated to the standard system with color terms and photometric zeropoints determined from sources in the field of SN~2021csp. Catalog magnitudes for local sequence stars were taken from the PanSTARRS catalog \citep{Chambers2016,Flewelling20} for $griz$, and from SDSS for $u$ \citep{Ahumada2020}. $B$ and $V$ magnitudes of the local sequence stars were converted from PanSTARRS $g$ and $r$ using the transformation derived by R. Lupton\footnote{\url{http://www.sdss3.org/dr8/algorithms/sdssUBVRITransform.php}.} Checks were performed to ensure the validity of  the transformation against APASS sources in the field, which indicated that the two were consistent to within the uncertainties. 
For $BVgriz$ images taken after MJD 59295 (+37~d in the rest frame), templates were subtracted using the {\sc hotpants} code\footnote{\url{https://github.com/acbecker/hotpants/}} before photometry was performed. Photometric measurements of SN~2021csp are reported in Table \ref{tab:phot_sloan}.

Space-based photometry was obtained with the Neil Gehrels \textit{Swift} Observatory \citep{gehrels04},  beginning 2021 Feb. 12 (PI Schulze).  Data from the Ultraviolet/Optical Telescope (UVOT; \citealt{roming2005}) were obtained from NASA's High Energy Astrophysics Science Archive Research Center (HEASARC\footnote{\url{https://heasarc.gsfc.nasa.gov/cgi-bin/W3Browse/swift.pl}}) and the {\it Swift} Quicklook site\footnote{\url{https://swift.gsfc.nasa.gov/sdc/ql/}}.  Photometry was performed using a pipeline based on the {\it Swift} Optical Ultraviolet Supernova Archive (SOUSA; \citealp{brown2014}).  In doing so, we made use of the \citet{breeveld2011} zeropoints   and the sensitivity correction updated in  2020.\footnote{\url{https://heasarc.gsfc.nasa.gov/docs/heasarc/caldb/swift/docs/uvot/uvotcaldb\_throughput\_06.pdf}}.  To be consistent with the sensitivity calibration, a 5\arcsec\ aperture was used throughout the analysis to compute flux and sky-level measurements.  Observations from 2012 Aug. 5 and 7 were used to numerically subtract the coincidence-loss corrected count-rate contribution from the host galaxy in the UV images. \textit{Swift}+UVOT photometry of SN~2021csp is reported in Table \ref{tab:phot_swift}.

\subsection{X-rays}

We used the online {\it Swift} XRT analysis tool \citep{Evans2020} to combine the XRT observations of SN~2021csp taken between 2021 Feb. 12 and 28 (total on-source time of $\sim 4.3$~hr). No source was detected in the combined image, and so we estimate an upper limit on the transient flux. Using the Bayesian methodology outlined by \cite{Kraft1991}, we calculate a 95\% confidence upper limit to the count rate of $4.2\times10^{-4}$~counts~s$^{-1}$. Assuming a neutral H column density of $2.37\times 10^{20}$~cm$^{-2}$ \citep{HI4PI2016} and a power-law source with photon index 2, we find an upper limit to the unabsorbed flux for SN~2021csp of $<1.55 \times 10^{-14}$~erg~cm$^{-2}$~s$^{-1}$.

We also checked Gamma-ray Coordination Network (GCN) alerts for any reported GRBs around the time of discovery of SN~2021csp. The {\it Integral} satellite saw a number of possible GRBs in the 48~hr period prior to the first detection, at significance between $3.6\sigma$ and $5.7\sigma$\footnote{\url{www.isdc.unige.ch/integral/science/grb}} \citep{Mereghetti03}. However, as these were detected by the anticoincidence shield, there is no localization, and it is hence impossible to confidently associate any GRBs with SN~2021csp. 

\section{Spectroscopy}
\label{sec:spec_evol}

In order to determine the redshift of SN~2021csp, a spectrum of the underlying host galaxy was extracted from the GTC (+OSIRIS) spectrum taken on 2021 Feb. 19. Measurements of the narrow host-galaxy emission lines H$\alpha$, [\ion{N}{2}], and [\ion{S}{2}] provide a redshift $z=0.0830\pm0.001$ (where the uncertainty is dominated by the systematic uncertainty of $\sim$300~\kms\ arising from the unknown peculiar velocity). This corresponds to a distance modulus $\mu=37.77^{+0.04}_{-0.05}$~mag, for our adopted $\Lambda$CDM cosmology (Sec.~\ref{sec:intro}). Our highest signal-to-noise ratio (S/N) spectra show no sign of narrow interstellar \ion{Na}{1}~D absorption, and are quite blue, indicating that extinction due to dust in the host galaxy is probably negligible.

\begin{figure*}
\includegraphics[width=\textwidth]{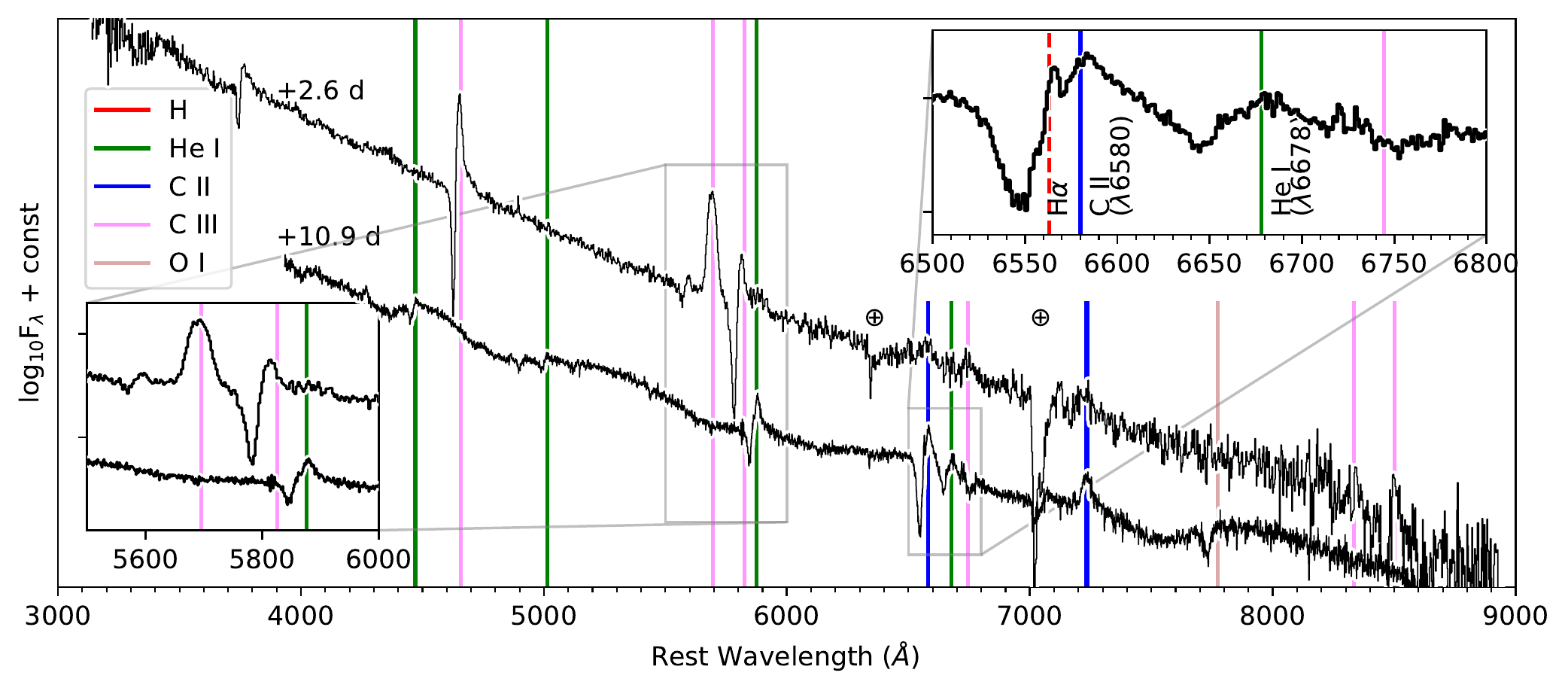}
\caption{Line identifications in our +2.6~d (NOT+ALFOSC) and +10.9~d (BAADE+IMACS) spectra of SN~2021csp. Vertical lines indicate the laboratory wavelengths of the stated ions. During the first week the spectra are dominated by narrow \ion{C}{3} features. However, as the object expands and cools the \ion{C}{3} gives way to \ion{C}{2}, \ion{He}{1}, and \ion{O}{1}. The unresolved H emission marked in the upper-right inset with a dashed red line is likely produced by underlying host-galaxy contamination.}
 \label{fig:lineID}
\end{figure*}

Fig.~\ref{fig:spec_sequence} shows the spectroscopic time series of SN~2021csp extending from $+$2.6~d to $+$51.8~d. Overall the continua of the early-time spectra are extremely blue, and are characterized by a number of prominent narrow features, consistent with the redshift of the host galaxy. During the first week, the spectrum is dominated by \ion{C}{3} lines. These features are identified and labeled  in Fig.~\ref{fig:lineID} and include   \ion{C}{3} $\lambda$4657, $\lambda$5696, $\lambda$5826, $\lambda$6744, $\lambda$8333, and $\lambda$8500. The \ion{C}{3} $\lambda$4657 and $\lambda$5826 lines show a strong P~Cygni profile. A constant velocity of $-v_{\rm abs} = 1800\pm100$~\kms\ is inferred from the position of the maximum absorption of the $\lambda$5826 feature. One line at $\sim 3765$~\AA\ remains unidentified in the +2.6~d spectrum. A \ion{He}{1} line is close at $\lambda$3769, but we see none of the other strong expected He lines at this phase. We find no \ion{C}{3} lines listed in the National Institute of Standards and Technology 
Atomic Spectra Database \citep{NIST} close to this wavelength, and while there are a number of other possibilities (namely Ca, Mg, O, and C), none of them are a strong candidate based on their transition strengths. 

While the \ion{C}{3} lines are initially strong, they  rapidly fade over the first week of evolution, and by +7.5~d are no longer present. However, during this period lines attributed to \ion{C}{2} $\lambda$6590, \ion{C}{2} $\lambda$7234, and \ion{He}{1} $\lambda\lambda$4471, 4921, 5016, 5876, and 6678 emerge (Fig. \ref{fig:lineID}). In addition, we see a narrow absorption feature that we associate with \ion{O}{1} $\lambda$7774, at the same velocity as the rest of the narrow CSM features.
By $+$7.5~d we also see the appearance of underlying bumps that hint toward broad SN-like features. A weak, narrow H$\alpha$ line is also discernible in emission. However, as this line is unresolved (compared to the other narrow lines with velocity of order 1000--2000~\kms), it is likely due to host-galaxy contamination rather than being intrinsic to the transient.

\begin{figure*}
\includegraphics[width=\textwidth]{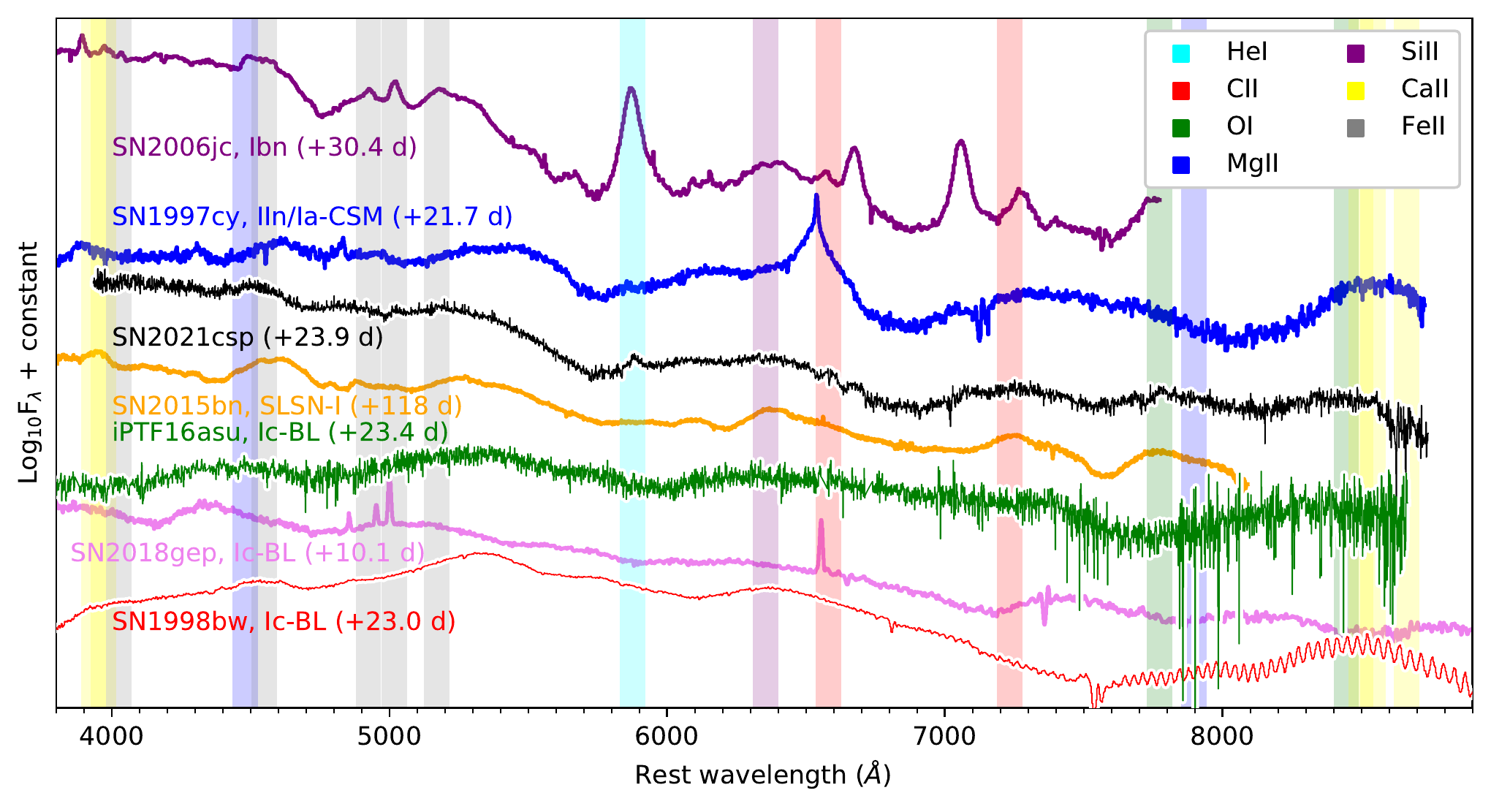}
\caption{Spectral comparison of the underlying SN of SN~2021csp compared with similar-phase spectra of SN~1998bw \citep{Patat2001}, iPTF16asu \citep{prictchard2020}, SN~2006jc \citep{Pastorello2007}, SN~2018gep \citep{prictchard2020}, and SN~1997cy \citep{Turatto00}, as well as a late-phase spectrum of SN~2015bn \citep{Nicholl2016}. Reported phases are in rest-frame days with respect to the discovery epoch. Prominent features are indicated with vertical colored bars and labeled.
} \label{fig:spec_comp_broad}
\end{figure*}

These narrow lines decrease in strength over the first two weeks from discovery. By $+$14.5~d they have disappeared, while broad undulations emerge and strengthen over time (Fig. \ref{fig:spec_sequence}), revealing the presence of an underlying SN. The SN displays no signs of H, and can be spectroscopically classified at this stage as Type Ib/c. We used the {\sc gelato} code\footnote{\url{gelato.tng.iac.es/}} \citep{Gelato08} to compare the +23.9~d spectrum of SN~2021csp to a library of templates. While there are no close matches, we find that both SNe~Ic-BL (such as SN~1998bw and iPTF16asu; \citealp{Patat2001,Pritchard20}) and the superluminous Type Ic SN~2015bn \citep{Nicholl2016} are quite similar (Fig. \ref{fig:spec_comp_broad}), providing support for the classification of the underlying SN as a Type Ic.

The spectral similarity to iPTF16asu is intriguing. This SN was a Type Ic-BL \citep{Whitesides17,Wang19}, which also displayed a fast rise time. The early-time spectra of iPTF16asu were blue and featureless, yet it is possible that narrow lines were present but not observed owing to the relatively low S/N of these spectra. Both \cite{Whitesides17} and \cite{Wang19} suggested that a central engine (possibly the spin-down of a young magnetar), as well as CSM interaction, was required to reproduce the light curve of iPTF16asu.

At +23.9~d, the spectrum exhibits a strong blue component below $\sim 5000$~\AA\ (indicated in Fig. \ref{fig:spec_sequence}). This is most likely emission from a forest of Fe lines, which form a pseudocontinuum. A similar pseudocontinuum has been seen in a number of other types of interacting SNe, including SNe~Ia-CSM \citep{Silverman13}, SNe~Ibn such as SN 2006jc \citep{Pastorello2007}, SNe~IIn
\citep{Stritzinger12}, and even the SN~Ic with late-time interaction SN~2017dio \citep{Kuncarayakti18}. A comparison to some of these SNe is shown in Fig. \ref{fig:spec_comp_broad}.

Interestingly, we also find a good match to the peculiar SN~1997cy \citep{Germany2000,Turatto00}. Aside from the narrow H$\alpha$ emission seen in SN~1997cy, the broad features appear to match well with SN~2021csp. The nature of SN~1997cy is debated; while it was originally called a very energetic explosion dominated by CSM interaction \citep{Germany2000,Turatto00,Nomoto1999}, more recently some authors have suggested it to be a SN~Ia-CSM \citep{Inserra16}, where a thermonuclear SN exploded inside a H-rich CSM likely produced by a binary companion. Within this paradigm, the apparent similarity to SN~2021csp may simply reflect that any SN explosion within a dense CSM may appear similar.

Between +18.4~d and +51.8~d we see relatively few changes in the spectra, the only significant one being the strengthening of the \ion{Ca}{2} triplet at $\lambda\lambda\lambda$8498, 8542, 8662 seen in Fig. \ref{fig:spec_sequence}. By the time of the final spectrum at +51.8~d, SN~2021csp does not appear to be optically thin, which is surprising for a normal SESN at this phase, although less unusual for a SN~Ic-BL.

Measuring the velocities seen in the broad lines is difficult, since the lines are shallow and it is difficult to unambiguously associate them with a single species. We opt to fit the Ca~NIR triplet in our +50.5~d spectrum using three Gaussian emission components fixed to their rest wavelengths, and with a common full width at half-maximum intensity (FWHM). The composite model provides a reasonable match to the observed spectrum for a FWHM of 6250~\kms. Such a velocity is not exceptional for an SESN at this phase \citep[e.g.,][]{Valenti2011}.

\begin{figure}
\includegraphics[width=\columnwidth]{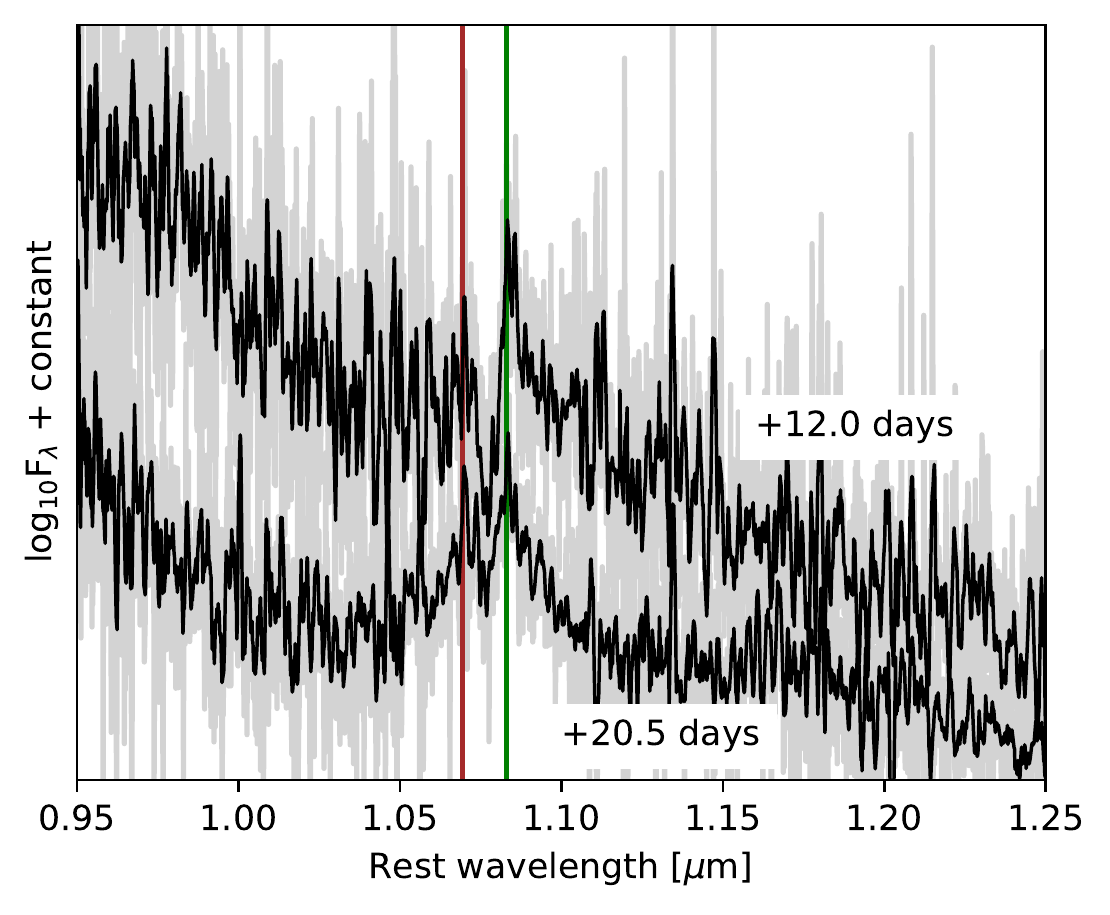}
\caption{NIR spectra of SN~2021csp. Gray lines show the original spectrum, black lines are after smoothing with a Savitzky-Golay filter across a 15~pixel window. The green vertical line marks the rest wavelength of the \ion{He}{1}~1.083~$\mu$m line, the brown line is \ion{C}{1}~1.0693~$\mu$m.}%
\label{fig:nir_spec}
\end{figure}

Unfortunately, our NIR spectra have relatively low S/N. No strong features are seen, but after smoothing and binning these data we detect narrow \ion{He}{1} at 1.083~$\mu$m and \ion{C}{1} 1.069~$\mu$m (Fig. \ref{fig:nir_spec}). The He line has a narrow P~Cygni profile with a minimum at 1800~\kms, consistent with the contemporaneous optical spectra.

\section{Photometric evolution} 
\label{sec:phot_evol}

\begin{figure*}
\includegraphics[width=1\textwidth]{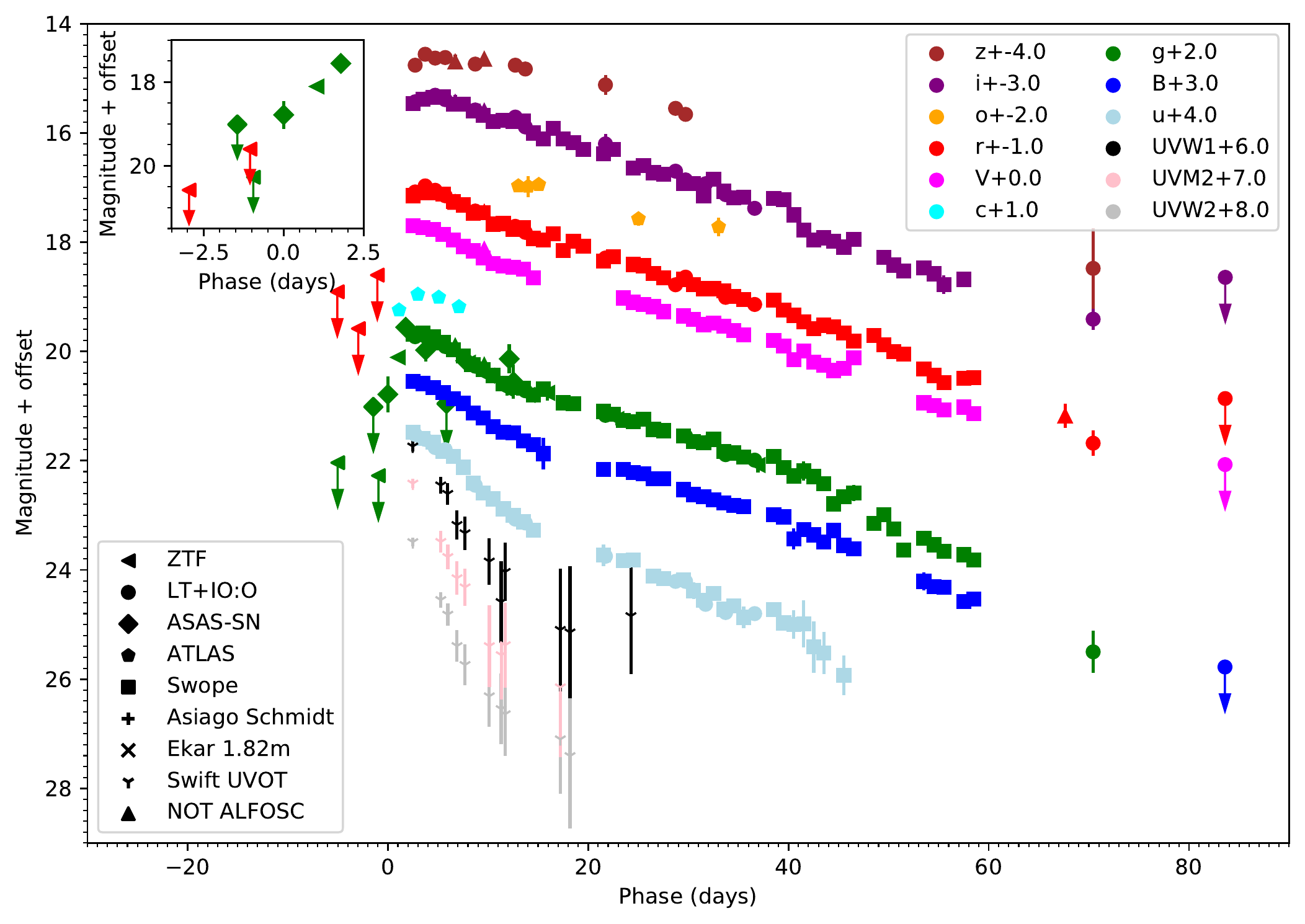}
\caption{Observed optical and UV broad-band light curves of SN~2021csp. For clarity, the light curves are offset by a constant as indicated in the upper legend. Limiting magnitudes are indicated with arrows, and the inset in the upper left shows the period around discovery, where we have highly constraining prediscovery limits.}
\label{fig:phot}
\end{figure*}

SN~2021csp has an exceptionally rapid rise to peak brightness, which is seen in the combined ASAS-SN and ZTF light curve (Fig. \ref{fig:phot}). Our first detection of SN~2021csp from ASAS-SN is at $g=18.79\pm0.33$~mag (on MJD=59255.47), with a restrictive nondetection from ZTF the previous night of $g>20.3$~mag (at $-0.9$~d). Two nights after discovery, SN~2021csp reaches its peak of $g=17.56\pm0.11$~mag (at +1.8~d). The rise time of SN~2021csp is hence between $\sim 2.7$~d and $\sim 1.8$~d, where the former is a hard limit from the ZTF nondetection. 

SN~2021csp reaches a peak absolute magnitude of $M_g = -20.3 \pm 0.1$~mag, significantly brighter than most SESNe. 
The fast rise seen in SN~2021csp to a bright peak is reminiscent of the evolution of SNe~Ibn \citep{Hosseinzadeh17}. In the case of SNe~Ibn, the fast rise is explained by a rapid increase in luminosity as the SN ejecta collide with CSM, although we note that for other SNe a fast rise may be due to a small ejecta mass and a short diffusion timescale. We use modeling to estimate the CSM configuration necessary to produce such a rapid rise in Sec.~\ref{sect:bolom_model}.

\begin{figure}
\includegraphics[width=\columnwidth]{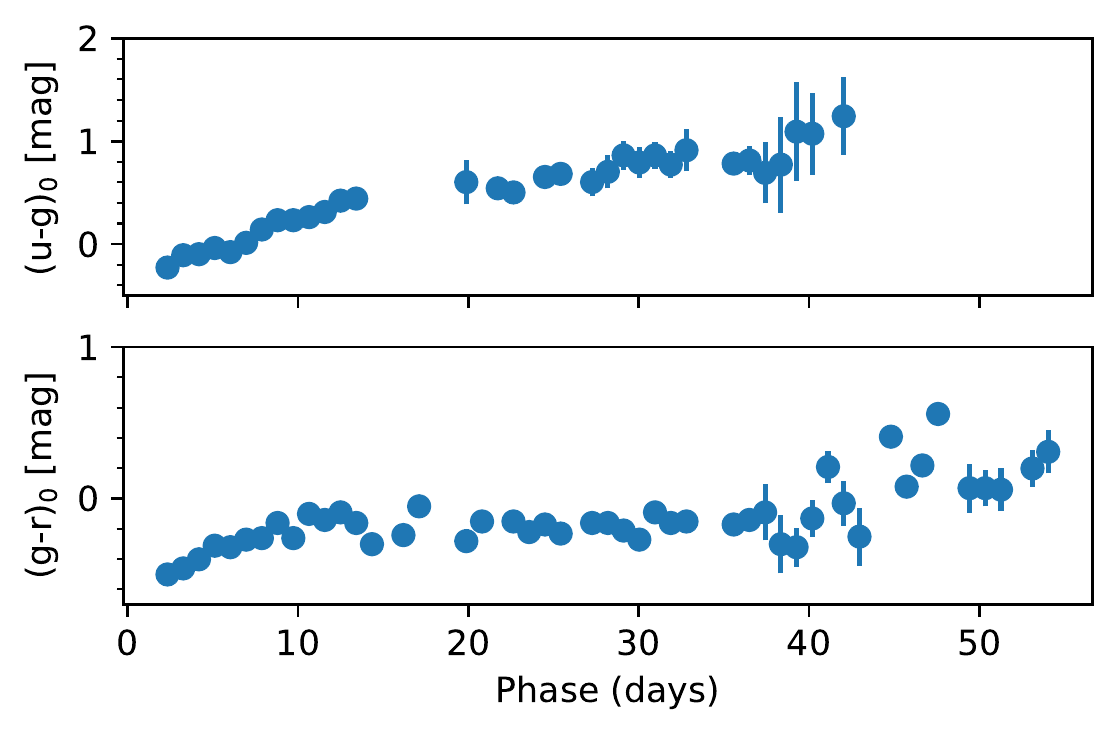}
\caption{Extinction-corrected color-curve evolution of SN~2021csp.}
\label{fig:colors}
\end{figure}

After maximum light, SN~2021csp displays a linear (in magnitudes) decline in all bands, with a slope of $0.058 \pm 0.001$~mag~day$^{-1}$ in $r$. This decline continues until the SN became too faint to follow, around two months after discovery. The bluer filters display a faster decline, most notable in the {\it Swift} UV filters (Fig. \ref{fig:phot}), and this can also be seen in the color evolution. In Fig.~\ref{fig:colors} we show the intrinsic color evolution of SN~2021csp over the first month after explosion. The $u-g$ color shown  becomes steadily redder for one month, while the $g-r$ color only evolves redder during the first week and then remains approximately constant.

\begin{figure}
\includegraphics[width=\columnwidth]{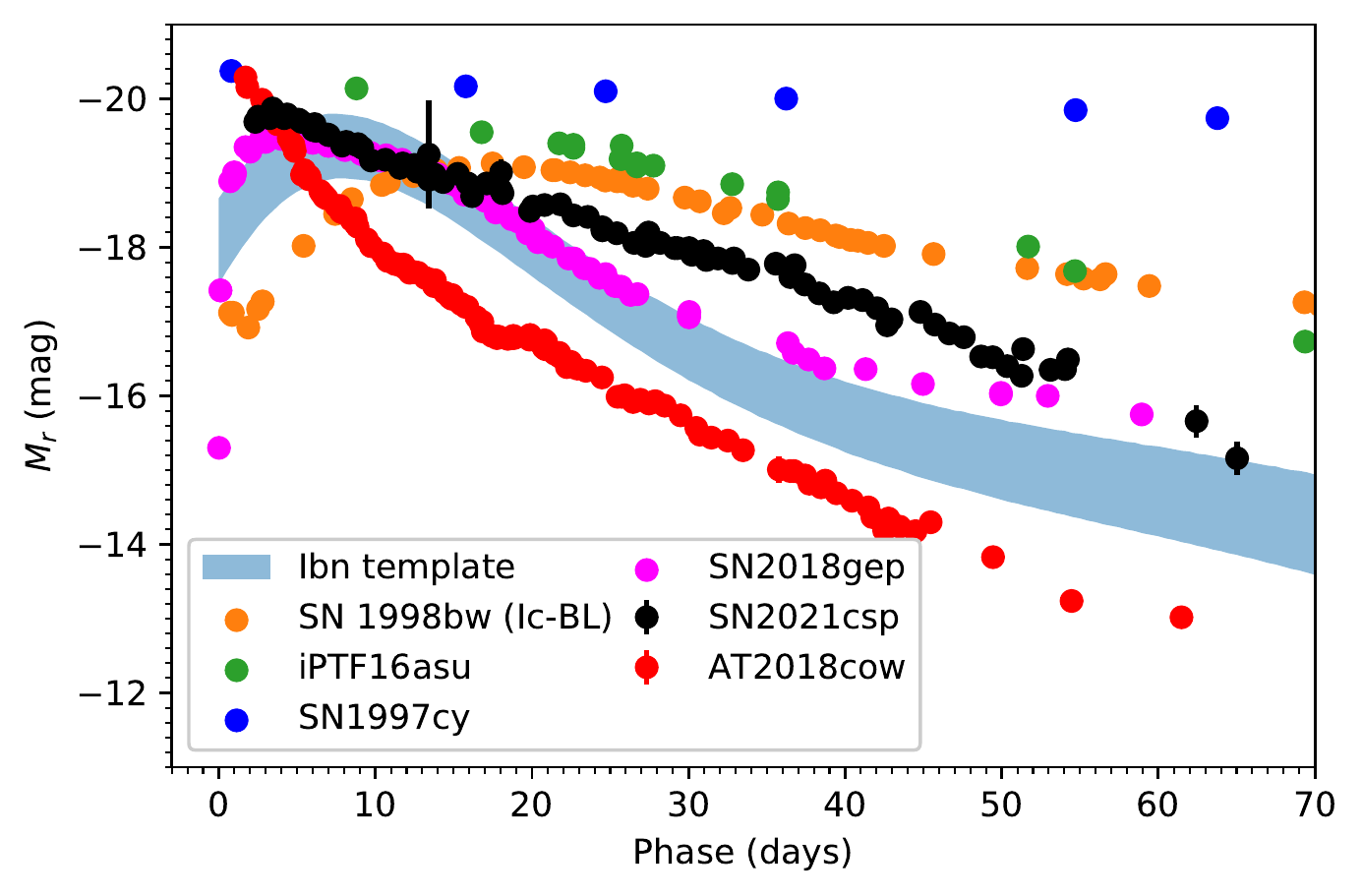}
\caption{Absolute $r$-band light curve of SN~2021csp compared to the $R$/$r$ light curve of a sample of FBOTs, SNe~Ic-BL, and other observationally similar or potentially related transients from the literature: AT~2018cow \citep{Prentice18}, 
SN~1997cy \citep{Germany2000}, a SN~Ibn template \citep{Hosseinzadeh17}, iPTF16asu \citep{Whitesides17}, SN~2018gep \citep{Ho19}, and SN~1998bw \citep{Patat2001}. Phase is with respect to the explosion epoch, or where this was unavailable, from discovery (we note that SN~1997cy has a highly uncertain explosion epoch).}
\label{fig:phot_comparison}
\end{figure}

We compare the light curve of SN~2021csp to that of several transients in Fig.~\ref{fig:phot_comparison}. As SN~2021csp is a (thus far) unique SN, we compare to a wide set of transients that are either spectroscopically similar at late times, exhibit a blue and largely featureless spectrum for a long period, or display signs of interaction with a H-poor CSM. We compare to the SN~Ibn template from \cite{Hosseinzadeh17} in Fig.~\ref{fig:phot_comparison}, as well as to the peculiar ``fast blue optical transient'' (FBOT) AT~2018cow \citep{Prentice18, Perley19}. We also compare to a set of SNe~Ic-BL (SN~1998bw, iPTF16asu, SN~2018gep; \citealp{Patat2001,Whitesides17,Ho19}) which are spectroscopically similar to SN~2021csp.

As expected, the peculiar FBOT AT~2018cow \citep{Prentice18} is initially quite similar (viz. bright absolute magnitude, fast rise, and a blue spectrum), but subsequently fades much more rapidly than SN~2021csp. On the other hand, SN~1997cy, which has some spectroscopic resemblance (Sec.~\ref{sec:spec_evol}), barely fades over two months.

SN~2021csp displays a similar photometric evolution to that of iPTF16asu and SN~2018gep, which all peak at $r \approx -20$~mag and have similar decline rates. Interestingly, while SN~2021csp is initially brighter than the Type Ic-BL SN~1998bw, after $\sim 2$ weeks the latter is brighter than the former. This reversal happens at the same phase as the narrow, presumably CSM interaction-dominated features disappear in SN~2021csp, and the broad underlying SN features emerge.

\section{Bolometric evolution and modeling}
\label{sect:bolom_model}

\begin{figure}
\includegraphics[trim={1.5cm 5cm 1.5cm 3cm},width=1\columnwidth]{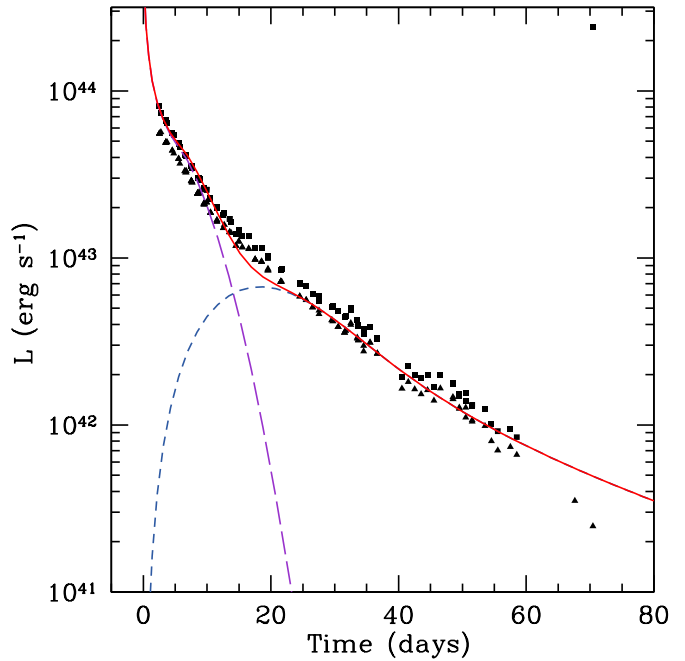}
\caption{The pseudobolometric (UVOT $UVW2$ to $z$; triangles) and bolometric (squares) light curve of SN~2021csp, compared to the model discussed in Sec.~\ref{sect:bolom_model} ($M_{\rm ej}=2.0$~\msun, $M_{\rm Ni}=0.4$~\msun, $E_{\rm SN} = 4\times10^{51}$~erg, $M_{\rm e}=1.0$~\msun, $R_{\rm e} = 400$~\rsun). The purple dashed line shows the shock-cooling emission from the CSM, while the blue dashed line indicates the luminosity from radioactive decay. The solid line shows the combined model luminosity, which well reproduces the observed bolometric light curve.}
\label{fig:arnett}
\end{figure}

We calculate a bolometric light curve of SN~2021csp from our optical and UV photometry using the {\sc superbol} code \citep{Nicholl_superbol}. The bolometric light curve is determined in two ways. First, we calculate a pseudobolometric light curve, integrating only over the observed wavelength range (i.e., from the {\it Swift}+UVOT UV bands to $z$). Second, we estimate a fully bolometric light curve, fitting a blackbody function to the spectral energy distribution (SED) at each epoch, and using this to extrapolate blueward and redward where we do not have data. Both light curves are shown in Fig.~\ref{fig:arnett}. SN~2021csp follows a nearly linear decline for around two months, with no sign of either a diffusion peak or a light-curve break where the SN settles onto a radioactively powered tail phase.

\begin{figure}
\includegraphics[width=\columnwidth]{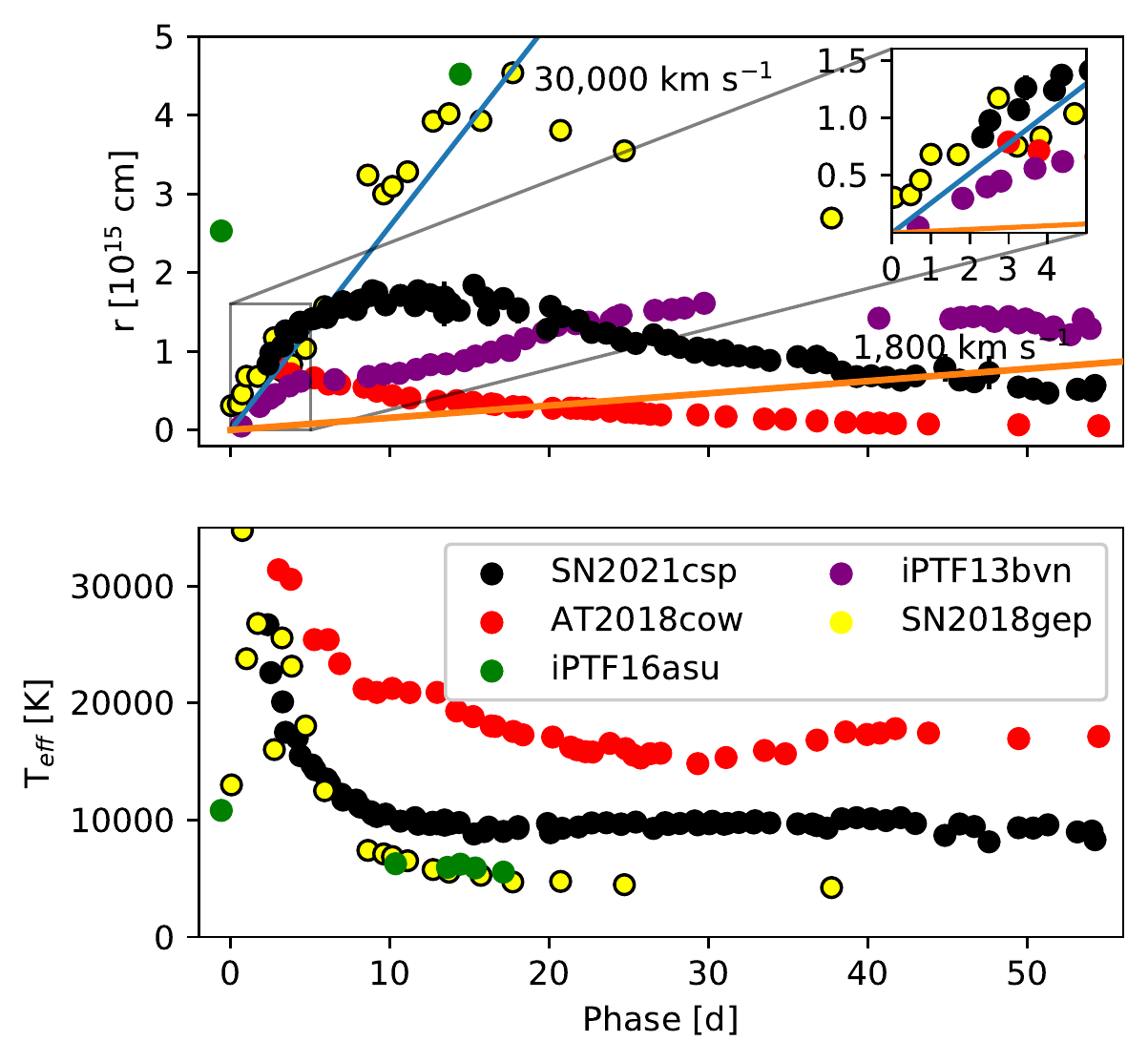}
\caption{Blackbody radius (upper panel) and effective-temperature evolution (lower panel) from SED fitting to SN~2021csp. Also shown are the radius and temperature for iPTF16asu \citep{Whitesides17}, AT~2018cow \citep{Perley19}, and SN~2018gep \citep{Ho19}, as well as the radius evolution of a normal SESN, iPTF13bvn \citep{Fremling16}. In the upper panel the radius evolution for a photosphere expanding at 30,000~\kms\ (blue line) and 1800~\kms\ (orange line) is shown, as well as a close-up view of the radius evolution over the first five days. Phase is with respect to explosion epoch (or in the case of AT2018cow, discovery).} \label{fig:bolom_params}
\end{figure}

Fitting the SED of SN~2021csp also allows us to infer a blackbody radius and temperature, which we plot in Fig. \ref{fig:bolom_params}. SNe deviate from true blackbodies due to wavelength-dependent opacity, limb darkening and the effect of broad spectral lines. However, blackbody fits can provide a useful guide to physical parameters, modulo effects such as incomplete gamma-ray trapping.

The blackbody radius of SN~2021csp is already $(0.83\pm0.08)\times10^{15}$~cm at +2.52~d (see inset in Fig.~\ref{fig:bolom_params}; the uncertainty here is the statistical error on the fit). If we assume a compact ($\lesssim 30$~\rsun) progenitor as would be typically expected for an SESN \citep{Yoon10}, this naively implies that the ejecta must have been expanding at $\sim 38,000$~\kms\ to reach this radius. This expansion velocity is of course sensitive to the adopted explosion epoch; however, even if we assume that SN~2021csp exploded immediately after the last ZTF nondetection, then it must still be expanding at $\sim 21,000$~\kms. Alternatively, and perhaps more plausibly, this blackbody radius could be accounted for by a pre-existing CSM, which is photoionized by the shock-breakout flash associated with the SN explosion.

In Fig.~\ref{fig:bolom_params} we compare the radius evolution of SN~2021csp to that of the Type Ic-BL SN~2018gep and iPTF16asu, as well as the normal SN~Ib iPTF13bvn \citep{Fremling16}. The radius of SN~2021csp expands rapidly at $\sim 30,000$~\kms\ over the first week after explosion. This is much faster than the 1800~\kms\ P~Cygni absorption seen in the spectra at this phase. The expansion is also faster than what is typically measured in an SESN (e.g., iPTF13bvn in Fig. \ref{fig:bolom_params}), although comparable to the blackbody velocity of SN~2016gep and iPTF16asu.
After 10 days, the photospheric radius starts to recede inward again, as is typically seen in SNe as the ejecta density and optical depth decrease \citep[e.g.][]{Ergon14}. However, the maximum radius of SN~2021csp is smaller than that of the SN~Ic-BL comparisons, and starts to decrease much earlier than that of iPTF13bvn.

We also see from Fig.~\ref{fig:bolom_params} that the temperature of SN~2021csp cools rapidly over the first week of evolution, before settling at $\sim 10,000$~K for the remainder of our observations. This is consistent with the color curves for SN~2021csp (Fig.~\ref{fig:colors}), in which we see a roughly constant $g-r=0$ color from 10 to 40 days after explosion. We note that this temperature is broadly consistent with the recombination temperature for He ($\sim 10,000$~K), while C and O recombine at a cooler temperature (6000~K). One must caution that the SEDs of SNe are often not well reproduced by a blackbody, especially at later phases. However, while broad emission lines and the pseudocontinuum in the blue undoubtedly affect the fit, we find a similar temperature (8200~K) from fitting our +50.5~d GTC (+ OSIRIS) spectrum.

To better understand the light curve observed for SN~2021csp, we attempt to fit it with a model that combines both circumstellar and nickel-powered components. The relatively smooth evolution of the light curve makes it difficult to fit these components uniquely without strong degeneracies. Nevertheless, some basic arguments can be used to get a rough idea of the parameters that reasonably describe this event.

The first important clue is the transition of the spectra at $\sim 20$~d from being hot and mostly featureless as expected for shock-cooling emission (SCE) to looking more like a typical SN~Ic-BL. If we propose that this is roughly the peak of what is a typical SN~Ic-BL powered by radioactive nickel ($^{56}$Ni) decay, using Arnett's Rule \citep{Arnett82} we infer an amount of synthesized nickel of $M_{\rm Ni} \approx 0.4~{\rm M}_\odot$. Although this provides a luminosity that roughly matches SN~2021csp at peak, it greatly overshoots the late-time tail. This is due to gamma-ray leakage, which can be used to provide better constraints on the ejecta mass $M_{\rm ej}$ and SN explosion energy $E_{\rm SN}$.

To estimate a time-dependent light curve due to the radioactive component, we use a one-dimensional model that solves a single differential equation. This was tested and found to closely match the work by \citet{Valenti08} and \citet{Lyman2016}, but with the added benefit that it smoothly transitions into times when gamma-ray leakage is strong. Diffusion through the ejecta is controlled by the diffusion timescale, defined as
\begin{eqnarray}
	\tau = \left( \frac{\kappa_{\rm opt}}{\beta c}\right)^{1/2}
		\left( \frac{6}{5}\frac{M_{\rm ej}^3}{E_{\rm SN}} \right)^{1/4},
\end{eqnarray}
where $\kappa_{\rm opt}$ is the optical opacity (taken to be $0.1~{\rm cm^2~g^{-1}}$ here) and $\beta=13.8$ is an eigenvalue from the diffusion problem \citep{Arnett82}. The time-dependent internal energy $E$ of the SN is described by
\begin{eqnarray}
	\frac{1}{t}\frac{d(Et)}{dt}
		= L_{\rm nuc}\left[ 1-\exp\left( -T_0^2/t^2\right)\right] - L_{\rm rad},
		\label{eq:diffusion}
\end{eqnarray}
where $L_{\rm nuc}$ is the typical time-dependent heating rate from the decay of $^{56}$Ni and subsequent $^{56}$Co, the radiative luminosity is
\begin{eqnarray}
    L_{\rm rad} = \frac{2tE}{\tau^2},
    \label{eq:lrad}
\end{eqnarray}
and
\begin{eqnarray}
    T_0 = \left( \frac{0.05\kappa_\gamma M_{\rm ej}^2}{E_{\rm SN}}\right)^{1/2},
\end{eqnarray}
is the diffusion timescale for gamma-rays \citep{Wheeler15}, where $\kappa_\gamma=0.03~{\rm cm^2~g^{-1}}$ is the gamma-ray opacity and the factor of $0.05$ is weakly dependent on the density profile of the ejecta.

For a chosen $M_{\rm Ni}$, $M_{\rm ej}$, and $E_{\rm SN}$, we integrate Eq.~(\ref{eq:diffusion}) forward in time to find $E(t)$ and then use Eq.~(\ref{eq:lrad}) to solve for $L_{\rm rad}(t)$ for comparison with SN~2021csp. In Fig.~\ref{fig:arnett}, we plot an example solution using $M_{\rm Ni}=0.4$~\msun, $M_{\rm ej}=2$~\msun, and $E_{\rm SN}=4\times10^{51}~{\rm erg}$ (blue short-dashed line). Although this implies that a significant fraction of the ejecta is $^{56}$Ni, these general parameters are similar to what \citet{Lyman2016} infer for the Type Ic-BL SN~2003jd and SN~2007ru. Also, note that there can be errors in the ejecta mass or nickel mass estimates when such simplistic models are used \citep{Khatami19} or recombination is not accounted for \citep{Piro14}, but this comparison at least shows that the underlying radioactively powered event in SN~2021csp is not dissimilar from other SNe~Ic-BL.

Next, we consider the earlier component ($<20$~d) powered by the SCE of dense CSM. For this, we use the analytic framework of \citet{Piro21}. In this model, the CSM is sufficiently dense that the SN shock propagates from the SN ejecta into the CSM, and only breaks out once it reaches the outer edge of the CSM. This introduces two additional parameters, the mass of the extended CSM $M_{\rm e}$ and its radius $R_{\rm e}$. The kinetic energy of the dense CSM is set by shock propagation and reflection at the boundary between the SN and CSM, as described by \citet{Nakar14}.

In Fig.~\ref{fig:arnett}, we plot an example SCE solution (purple long-dashed line) as well as the sum of the SCE and radioactively powered components (red solid line). The excess early emission from SN~2021csp lasts $\gtrsim15$~d, which is much longer than, for example, the early radiative bump seen from many SNe~IIb (roughly a few days). Thus, the dense extended material must be much more massive. For this specific example, we use $M_{\rm e} \approx 1$~M$_\odot$ and $R_{\rm e} \approx 400$~R$_\odot$. We note that there is some degeneracy between $M_{\rm e}$ and $R_{\rm e}$ since the early SCE luminosity scales $\propto R_{\rm e}/M_{\rm e}$, although $M_{\rm e}$ must be sufficiently large that the extended material does not become optically thin too soon and the SCE drop off too quickly.

The origin of this massive, extended material that causes the bright SCE provides important clues about the ways in which massive stars end their lives. The scale is similar to that of SNe~IIb, which also often show prominent (albeit shorter lived) SCE. A common aspect between these events is the presence of a binary companion, so it is natural to think that binary interactions are playing a role in generating the dense CSM. The mass of the CSM is a factor of $\sim 10$--100 larger for SN~2021csp, perhaps not surprising given that the lack of H or strong He features suggests binary stripping must have been much stronger in this case.

Alternatively, the CSM may be created by a pre-SN outburst \citep[e.g.,][]{Fraser13}. While such outbursts have been retrospectively identified in archival survey data for other interacting SNe, the distance of SN~2021csp is such that even relatively deep ZTF images would only be sensitive to outbursts with magnitudes typical of bright supernovae near maximum light.

\section{Discussion and conclusions}
\label{sect:nature}

We find evidence that SN~2021csp interacts with an H-free and He-poor CSM over the first 10~d of its evolution. This CSM is fast, with a velocity of 1800~\kms, consistent with that seen in Wolf-Rayet stars \citep[e.g.][]{Crowther07}. After two weeks, the SN has evolved to resemble a SN~Ic-BL, although with a pronounced pseudocontinuum in the blue that is indicative of ongoing CSM interaction.

Using a semi-analytic model that includes shock-cooling emission from an extended CSM, as well as an Arnett-like diffusion model for radioactive decay, we find that the bolometric light curve can be reproduced by a $4\times10^{51}$~erg explosion with 2~\msun\ of ejecta and 0.4~\msun\ of $^{56}$Ni, plus a contribution from SCE of $\sim 1$~\msun\ of CSM extending out to 400~\rsun.
These model parameters are consistent with what is found for SNe~Ic-BL, although we note that around 20\% of the ejecta must be $^{56}$Ni, which is perhaps a little high. However, it is possible that a higher trapping of gamma-rays and positrons may allow for a solution with a smaller $^{56}$Ni fraction in the ejecta.

SN~2021csp appears to be connected to the SNe~Ic-BL that show evidence for a significant CSM prior to explosion (e.g., iPTF16asu, SN2018gep; \citealp{Whitesides17,Ho19,Wang19}). On the one hand, the existence of such SNe does not seem surprising --- SNe~Ibc have long been suggested to arise from massive Wolf-Rayet stars, which show fast winds and significant mass loss \citep[see][for a review]{Crowther07}. On the other hand, the majority of SESNe do {\it not} show evidence for large amounts of CSM close by, indicating that there is something unusual about the progenitors of a subset of these SNe. Two potential avenues that would be interesting to explore are whether a specific binary progenitor with a particular mass ratio and initial period, or perhaps a geometric or viewing-angle effect, can reproduce the CSM seen in SN~2021csp.

However, recent radio observations have suggested that even the progenitors of more normal SN~Ic progenitors such as that of SN~2020oi may have experienced variations in mass-loss rate less than a year prior to explosion (\citealp{Maeda21}; although see also \citealp{Horesh20,Gagliano21}). Such variability is likely connected to instabilities during the final nuclear burning stages in the progenitor. The CSM for SN~2021csp is clearly larger than that around SN~2020oi; however, it is tempting to draw a connection between the two, where a more energetic pre-SN outburst or superwind phase produced a more massive CSM.

Turning to the spectra, the apparent similarity of SN~2021csp to SNe~Ic-BL is intriguing. So far, none of the purported SNe~Ic-BL have an associated GRB detection. This could be due to these SNe not launching a jet (or perhaps producing a failed GRB). On the other hand, it may be due to a combination of distance and an off-axis viewing angle. In the latter case, it is simply a matter of time and luck before one such event {\it is} detected in gamma-rays. As well as SNe~Ic-BL, SN~2021csp also shows a resemblance to SLSNe, albeit at a much later phase. This apparent similarity may be due to both SNe~Ic-BL and SLSNe being powered by a central engine \citep[e.g. as discussed in ][]{Margalit18}, and it raises the prospect that the mass loss seen in SN~2021csp may also have relevance for the progenitors of SLSNe.

SN~2021csp is one of the prototypes of the new class of SNe~Icn (\citealp{GalYamIcn} also propose that several other SNe fall into this class, namely SNe~2010mb, 2019hgp, and 2021ckj). SN taxonomy is based largely on observational rather than physical properties \citep[e.g.,][]{Filippenko1997,galyam2017}, and we note that if SN~2021csp was classified around one week after discovery it would be regarded as an SN~Ibn. If it had been first observed several weeks after discovery, it would be classified as a SN~Ic-BL. While such a situation may appear unsatisfactory, it is inevitable given our limited understanding of the physical nature of these SNe, their progenitors, and the connections between the different subtypes.

One promising avenue to explore in the future would be detailed modeling of the narrow emission lines seen in SN~2021csp over the first 10~d after explosion. Such modeling has already been applied to very early-time spectra of more normal core-collapse SNe \citep[e.g.,][]{Groh14}, where it can provide quantitative constraints on both the progenitor mass-loss rate and metallicity. From an observational perspective, SN~2021csp yet again illustrates the need for both high-cadence imaging surveys (to catch transients with a very fast rise), as well as rapid, high-S/N and moderate-resolution spectroscopy that enables the identification of similar narrow He and He-poor features that would otherwise be missed.

\acknowledgments

M.F. is supported by a Royal Society -- Science Foundation Ireland University Research Fellowship.
M.D.S. is supported by grants from the VILLUM FONDEN (grant number 28021) and by  the Independent Research Fund Denmark (IRFD; 8021-00170B).
Y.-Z.C. is funded by China Postdoctoral Science Foundation (grant 2021M691821).
S.B. is supported by a Royal Society -- Science Foundation Ireland Enhancement Award (RS-EA/3471).
N.E.R. acknowledges support from MIUR, PRIN 2017 (grant 20179ZF5KS).
Support for T.W.-S.H. was provided by NASA through the NASA Hubble Fellowship grant HST-HF2-51458.001-A awarded by the Space Telescope Science Institute, which is operated by the Association of Universities for Research in Astronomy, Inc., for NASA, under contract NAS5-26555.
L.G. acknowledges financial support from the Spanish Ministry of Science, Innovation and Universities (MICIU) under the 2019 Ram\'on y Cajal program RYC2019-027683 and from the Spanish MICIU project PID2020-115253GA-I00.
M.A.T. acknowledges support from the DOE CSGF through grant DE-SC0019323.
S.M. acknowledges support from the Magnus Ehrnrooth Foundation and the Vilho, Yrj\"o and Kalle V\"ais\"al\"a Foundation.
A.R. acknowledges support from ANID BECAS/DOCTORADO NACIONAL 21202412.
P.H. acknowledges support from NSF grant AST-1715133.
NUTS2 access to the Nordic Optical Telescope (NOT) is funded partially by the Instrument Center for Danish Astrophysics (IDA).
A.V.F.'s group is supported by the Miller Institute for Basic Research in Science, the Christopher R. Redlich Fund, Gary \& Cynthia Bengier, Clark \& Sharon Winslow, Sanford Robinson, and many other individual donors.
X.-F.W. is supported by NSFC (grants 12033003 and
11633002), the Major State Basic Research Development
Program (grant 2016YFA0400803), and the Scholar Program
of Beijing Academy of Science and Technology (DZ: BS202002).
J.Z. is supported by the National Natural Science Foundation of China (NSFC, grants 11773067, 11403096), by the Youth Innovation Promotion Association of the CAS (grant 2018081), and by the Ten Thousand Talents Program of Yunnan for Top-notch Young Talents.

We thank the staffs of the various observatories where data were obtained for their assistance.

Based in part on observations made with the  NOT, owned in collaboration by  Aarhus University and the University of Turku, and operated jointly by Aarhus University, the University of Turku and the University of Oslo, representing Denmark, Finland and Norway, the University of Iceland and Stockholm University at the Observatorio del Roque de los Muchachos, La Palma, Spain, of the Instituto de Astrofisica de Canarias. The data presented here were obtained in part with ALFOSC, which is provided by the Instituto de Astrofisica de Andalucia (IAA) under a joint agreement with the University of Copenhagen and NOTSA.
Based in part on observations collected at Copernico and Schmidt telescopes (Asiago, Italy) of the INAF -- Osservatorio Astronomico di Padova.
Based in part on observations made with the Gran Telescopio Canarias (GTC), installed in the Spanish Observatorio del Roque de los Muchachos of the Instituto de AstrofÃ­sica de Canarias, in the island of La Palma (programme GTC9-20B).
Based in part on observations made with the Italian Telescopio Nazionale Galileo (TNG) operated on the island of La Palma by the Fundaci\'on Galileo Galilei of the INAF (Istituto Nazionale di Astrofisica) at the Spanish Observatorio del Roque de los Muchachos of the Instituto de Astrofisica de Canarias
Based in part on observations made with facilities at the Las Campanas Observatory including the Swope and Magellan telescopes.
We acknowledge the support of the staff of the LJT. Funding for the LJT has been provided by the Chinese Academy of Sciences and the People's Government of Yunnan Province. The LJT is jointly operated and administrated by YNAO and Center for Astronomical Mega-Science, CAS.

Keck NIRES data presented in the paper were supported by NASA Keck PI Data Awards 2020B\_N141 and 2021A\_N147 (JPL RSAs 1654540 and 1660953, PI S. W. Jha), administered by the NASA Exoplanet Science Institute. Some of the W.~M. Keck Observatory data presented herein were obtained from telescope time allocated to NASA through the agency's scientific partnership with the California Institute of Technology and the University of California. The Observatory was made possible by the generous financial support of the W.~M. Keck Foundation.
The authors wish to recognize and acknowledge the very significant cultural role and reverence that the summit of Maunakea has always had within the indigenous Hawaiian community. We are most fortunate to have the opportunity to conduct observations from this mountain.

This work has made use of data from the Asteroid Terrestrial-impact Last Alert System (ATLAS) project. The Asteroid Terrestrial-impact Last Alert System (ATLAS) project is primarily funded to search for near-Earth objects through NASA grants NN12AR55G, 80NSSC18K0284, and 80NSSC18K1575; byproducts of the NEO search include images and catalogs from the survey area. This work was partially funded by Kepler/K2 grant J1944/80NSSC19K0112 and HST GO-15889, and STFC grants ST/T000198/1 and ST/S006109/1. The ATLAS science products have been made possible through the contributions of the University of Hawaii Institute for Astronomy, the Queen's University Belfast, the Space Telescope Science Institute, the South African Astronomical Observatory, and The Millennium Institute of Astrophysics (MAS), Chile.

A major upgrade of the Kast spectrograph on the Shane 3~m telescope at Lick Observatory was made possible through generous gifts from the Heising-Simons Foundation as well as William and Marina Kast. Research at Lick Observatory is partially supported by a generous gift from Google.


%

\vspace{5mm}
\facilities{Asiago:Copernico(AFOSC),
    Asiago:Schmidt,
    NOT(ALFOSC and NOTCAM),
    Liverpool:2m(IO:O),
    Keck:I(LIRS), 
    Keck:II(NIRES), 
    Las Campanas Observatory:Swope,
    Swift(XRT and UVOT),
    Magellan:Baade(IMACS),
    Shane(Kast),
    UH:2.2m(SNIFS),
    GTC(OSIRIS),
    YAO:2.4m(YFOSC)}

%

\begin{deluxetable*}{ccclll}
\tablenum{1}
\tablecaption{Log of spectroscopic observations of SN~2021csp\label{tab:spec}}
\tablewidth{0pt}
\tablehead{
\colhead{Date} & \colhead{Phase} &  \colhead{Telescope} & \colhead{Instrument} & \colhead{Range} & \colhead{Resolution}\\
\colhead{UT} & \colhead{Days} &  \colhead{} & \colhead{} & \colhead{\AA} & \colhead{FWHM (\AA)}
}
\startdata
2021-02-13.35   & 2.6   & NOT               & ALFOSC+Gr4            & 3400--9680   & 14   \\ 
2021-02-13.82   & 3.1   & LJT               & YFOSC+G3              & 3500--7500   & 21   \\ 
2021-02-14.16   & 3.4   & Copernico 1.82m   & AFOSC+VPH7            & 3150--7280   & 15   \\ 
2021-02-14.88   & 4.1   & LJT               & YFOSC+G3              & 3500--7500   & 21   \\ 
2021-02-15.14   & 4.3   & Copernico 1.82m   & AFOSC+VPH7            & 3400--7280   & 14   \\ 
2021-02-16.07   & 5.2   & Copernico 1.82m   & AFOSC+gm4             & 3400--8170   & 13   \\ 
2021-02-17.23   & 6.2   & NOT               & ALFOSC+Gr4            & 3400--9650   & 19   \\ 
2021-02-18.66   & 7.5   & Shane             & Kast+600/300          & 3630--10340  & 9/12 \\ 
2021-02-19.19   & 7.9   & GTC               & OSIRIS+R1000B/R1000R  & 3620--10390  & 7/8  \\ 
2021-02-20.18   & 9.0   & NOT               & ALFOSC+Gr4            & 3400--9640   & 14   \\
2021-02-22.28   & 10.9  & Magellan Baade    & IMACS                 & 4260--9470   & 5    \\ 
2021-02-22.55   & 11.2  & UH88              & SNIFS                 & 3400--8080   & 3/7   \\ 
2021-02-23.43   & 12.0  & Keck-II           & NIRES+XD-110          & 9650--24670  & 6   \\ 
2021-02-24.90   & 13.3  & LJT               & YFOSC+G3              & 3500--8760   & 20   \\
2021-02-26.14   & 14.5  & NOT               & ALFOSC+Gr4            & 3400--9650   & 14   \\
2021-03-02.37   & 18.4  & Magellan Baade    & IMACS+Grism300/17.5   & 4260--9470   & 5    \\ 
2021-03-04.66   & 20.5  & Keck-II           & NIRES+XD-110          & 9650--24670  & 6    \\ 
2021-03-08.33   & 23.9  & Magellan Baade    & IMACS+Grism300/17.5   & 4260--9470   & 5    \\
2021-03-12.11   & 27.4  & NOT               & ALFOSC+Gr4            & 3400--9690   & 14   \\
2021-03-18.17   & 33.0  & TNG               & LRS+LRB/LRR           & 3350--10000  & 10/8 \\
2021-03-22.09   & 37.5  & NOT               & ALFOSC+Gr4            & 3400--9680   & 14   \\
2021-04-05.17   & 50.5  & GTC               & OSIRIS+R1000B/R1000R  & 3620--10400  & 7/8  \\ 
2021-04-07.60   & 51.8  & Keck-I            & LRIS+600/400          & 3130--10280  & 5/9  \\
\enddata
\tablecomments{Phase is listed in rest-frame days with respect to the ASAS-SN discovery epoch of MJD 59255.47.}
\end{deluxetable*}

\begin{deluxetable*}{cllllllllr}
\tablenum{2}
\tablecaption{Ground-based photometry of SN~2021csp. Sloan filters are reported in AB magnitudes, $BV$ filters are reported in Vega-based magnitudes. Uncertainties are given in parentheses. A portion of the table is shown here; the full table will be made available in electronic format online.\label{tab:phot_sloan}}
\tablewidth{0pt}
\tablehead{
\colhead{MJD} &  \colhead{$u$} & \colhead{$g$} & \colhead{$r$}  & \colhead{$i$}  & \colhead{$z$}  & \colhead{$B$}  & \colhead{$V$}  & \colhead{Instrument}
}
\startdata
59258.0 	& 17.48 (0.05) & 17.68 (0.01) & 18.15 (0.01) & 18.46 (0.01)	& -- 	       & 17.55 (0.06) &	17.70 (0.03) &	Swope \\
59258.2 	& 17.52 (0.05) & 17.73 (0.02) & 18.08 (0.02) & 18.43 (0.04)	& 18.76 (0.06) & 17.60 (0.04) & 17.71 (0.04) &	LT 	   \\
...         & ...          & ...          & ...          & ...          & ...          & ...          & ...          & ...    \\
 \enddata
\end{deluxetable*}

\begin{deluxetable*}{llllllll}
\tablenum{3}
\tablecaption{Swift+UVOT photometry for SN~2021csp. Uncertainties are given in parentheses. A portion of the table is shown here, the full table will be made available in electronic format online.\label{tab:phot_swift}}
\tablewidth{0pt}
\tablehead{
\colhead{MJD} & \colhead{$UVW2$} & \colhead{$UVM2$} & \colhead{$UVW1$}  & \colhead{$U$}  & \colhead{$B$}  & \colhead{$V$}  & 
}\startdata
59257.96      & 15.490 (0.076) & 15.419 (0.083) & 16.452 (0.151) & 16.236 (0.089) & 17.476 (0.128) & 17.610 (0.332)  \\
59260.75      & 16.550 (0.144) & 16.487 (0.199) & 16.452 (0.151) & 16.518 (0.105) & 17.743 (0.155) & 17.365 (0.289)  \\
...           & ...            & ...            & ...            & ...            & ...            & ...             \\
 \enddata
\end{deluxetable*}




\bibliographystyle{aasjournal}



\end{document}